\documentclass[11pt,dvips]{article}
cm \voffset = -2truecm

\usepackage{graphicx}
\usepackage{amssymb}
\usepackage{latexsym}

\begin{document}
\begin{titlepage}
\setcounter{page}{1}
\renewcommand{\thefootnote}{\fnsymbol{footnote}}

\vspace{5mm}
\begin{center}


 {\Large \bf Entangled Phase States via Quantum Beam Splitter}

\vspace{5mm}

{\bf M. Daoud$^{a,b}$\footnote {daoud@pks.mpg.de --
m$_-$daoud@hotmail.com}}
 and
{\bf E. B. Choubabi$^{c}$\footnote{choubabi@gmail.com}}

\vspace{5mm}

{$^{a}$\em  Max Planck Institute for Physics of Complex Systems,
 D-01187 Dresden, Germany}

{$^{b}$\em Department of Physics,
Faculty of Sciences, Ibn Zohr University},\\
{\em PO Box 8106,  80006 Agadir, Morocco}



{$^{c}$\em Theoretical Physics Group,  
Faculty of Sciences, Choua\"ib Doukkali University},\\
{\em PO Box 20,  24000 El Jadida, Morocco}

\vspace{3cm}

\begin{abstract}

We study the entanglement effect of beam splitter on the temporally
stable phase states. Specifically, we consider the eigenstates
(phase states) of an unitary phase operator resulting from the polar
decomposition of ladder operators of generalized Weyl--Heisenberg
algebras possessing finite dimensional representation space. The
linear entropy that measures the degree of entanglement at the
output of the beam splitter is analytically obtained. We find that
the entanglement is not only strongly dependent on the Hilbert space
dimension but also quite related to strength the parameter ensuring
the temporal stability of the phase states. Finally, we discuss the
evolution of the entangled phase states.

\end{abstract}

\end{center}

\end{titlepage}

\newpage

\section{Introduction}

Probably the entanglement phenomenon, often named quantum
non-locality \cite{Shro1,Shro2},
 contains one of the most interesting features of quantum
mechanics. Actually, it is at the heart of current development of
quantum information processing such as quantum teleportation
\cite{Ben1}, superdense coding \cite{Ben2} and telecloning
\cite{Murao}. Entanglement also plays a key role in secure
communication, for instance the Ekert protocol \cite{Eckert} based
on entangled states is more robust than  BB84 one \cite{BB84}.  It
known that in quantum
computation, 
the qubits are massively entangled. The preparation and
characterization of entangled optical as well as atomic states has
been studied extensively. In this respect and in a recent
experimental advance, the polarization-entangled photons were
generated using type I or type II parametric down conversion
\cite{Kwiat}.

The production of entangled states belonging to an infinite
dimensional Hilbert space can be achieved also by adopting the
standard technique of parametric down conversion \cite{Furusawa}.
The beam splitter is also one of the few experimentally accessible
devices,
 which may be used to generate entangled states. In this sense, different developments have been
 reported on the analysis 
of beam splitter as entangler
\cite{Tan,Sanders2,Paris,Kim,Markham,Gerry1}. In particular, the
effect of beam splitting on the spin (or SU(2)) coherent states for
a single mode field was investigated in \cite{Markham}. In the same
spirit the entanglement via a beam splitter of $SU(1,1)$ coherent
states (Barut-Giradello \cite{Barut} and Perelomov ones
\cite{Perelomov}) was discussed in \cite{Gerry1}. The investigation
of entanglement properties of coherent states \cite
{Sanders,Gerry2,Luis,Wang} is mainly motivated by the fact that  the
entangled nonorthogonal states also play an important role in the
quantum cryptography \cite{Fuchs} and quantum information processing
\cite{Jeong}. 
Experimentally, the quantum optical systems are extensively
investigated in order to generate, characterize and understand the
entanglement properties.

In this work, we focus on the analysis of the entangled phase states
of single modes of the electromagnetic field. It is well-known that
the usual way 
to quantize these single modes 
is through the harmonic oscillator techniques with an infinity of
states. In 1989, Pegg and Barnett~\cite{Pegg} suggested to truncate
up to some finite, but arbitrarily large, order the infinite
dimensional representation space of the harmonic oscillator algebra.
This was done to get  rid of the difficulty related to the infinite
dimensional character of the representation space,  which
constitutes a drawback in defining a phase operator in a consistent
way \cite{Louisell,Susskind,Carruthers}. Here we introduce a
generalized version of the  Weyl--Heisenberg algebra that allows us
to achieve our goal. We particularly consider one algebra possessing
finite dimensional bosonic representations. This is essential in
defining the Hermitian phase operator and  corresponding temporally
stable
phase states. For this purpose, we  use a technique based on 
an approach developed in~\cite{Daoud}. Furthermore, we show 
that the phase parameter ensuring the temporal stability of phase
states plays a crucial role in the present analysis. Finally, we
deal with 
the entanglement of phase states when passed trough a beam splitter.

The outline of the paper is as follows. In section 2, we introduce a
generalized Weyl--Heisenberg algebra, which extends the dynamical
symmetry of the usual harmonic oscillator. Subsequently, we discuss
the corresponding finite dimensional Hilbertian representation and
derive the temporally stable
phase states, which are obtained to be dependent of  $\varphi$ called phase parameter. 
This latter allows us to take into account the nature of the
spectrum
of the system. Because of the absence of $\varphi$, 
one can immediately notice that the $SU(2)$ phase states derived in
\cite{Vourdas}
are identical to those obtained by Pegg and Barnett~\cite{Pegg} for
truncated harmonic oscillator. Furthermore, we analyze the basic
features
of the phase states. 
In section 3, we examine the entanglement resulting from the action
of a beam splitter on the phase states.  To investigate the degree
of bipartite entanglement of the phase states, we determine the
linear entropy.  Finally, we close by  some concluding remarks.

\section{Finite Fock space for generalized  Weyl--Heisenberg algebra}

A basic ingredient that will be used in the forthcoming analysis is
the generalized Weyl--Heisenberg algebra. This is generated
algebraically by three elements denoted by $\{a^+ , a^- ,N \}$,
which are satisfying the commutation relations
\begin{eqnarray}\label{wha}
\left[N, a^-\right] = -~a^-,\qquad \left[N, a^+ \right] = +
~a^+,\qquad
 \left[a^- , a^+\right] =  G(N)
\end{eqnarray}
where $G(N) = \left[G(N)\right]^{\dagger}$ is a Hermitian function
of the number operator $N$. Clearly, by requiring that $G(N) = {\bf
I}$, where ${\bf I}$ is the unity operator, we end up with  the
usual harmonic oscillator algebra.  Also, for $G(N) = a N + b$, with
two real parameters $a$ and $b$,  (\ref{wha}) reproduces the $W_k$
algebra discussed
 in the context of fractional supersymmetric quantum mechanics~\cite{Daoud2}. It is also important to stress
that this algebra covers the extended harmonic oscillator worked out
in \cite{Quesne1,Quesne2}.

Let us consider the abstract Fock representation of the above
algebra through a complete set of orthonormal states $\{ \vert n
\rangle ,  n \in \mathbb{N}\}$ those are eigenstates of the number
operator $N$, $N\vert n \rangle =n \vert n \rangle$. In this
representation, the vacuum state defined as $a^- \vert 0 \rangle =0$
 and the ortho-normalized eigenstates
constructed  by successive applications of the creation operator
$a^+$. Indeed, we define the actions of creation and annihilation
operators as
\begin{eqnarray}
 a^-\vert n \rangle &=& \sqrt{F(n)}  e^{{ i [F(n) - F(n-1)]\varphi}} \vert n - 1 \rangle\nonumber\\ 
 a^+\vert n \rangle &=& \sqrt{F(n+1)} e^{{-i [F(n+1)- F(n)]\varphi}} \vert n+1  \rangle
\label{action sur les n}
\end{eqnarray}
where the structure function $F(.)$ is an analytic function, with
the properties $F(0)=0$ and $F(n)>0$ for $n=1,2,\cdots$. The phase
parameter $\varphi$ will be discussed in the next by emphasizing its
play a crucial role in constructing the phase states. In what
follows we shall denote the Fock space  as ${\cal F}$ where the
operators $a^+$ and $a^-$ are mutually adjoint, $a^+ =
(a^-)^{\dagger}$ on ${\cal F}$. It is easy to check that  $F(.)$
satisfies the
 recursion relation
\begin{eqnarray}
 F(n+1) - F(n) = G(n)
\end{eqnarray}
which gives by simple iteration the form
\begin{eqnarray}
F(n) = \sum_{m=0}^{n-1} G(m). \label{sum}
\end{eqnarray}

In the forthcoming analysis, we restrict ourselves to generalized
oscillator algebra defined through  the structure functions those
fulfilling the condition
\begin{eqnarray}
 F(2s+1)=0
\label{condition}
\end{eqnarray}
where $2s$ is a positive integer value. It this case, it follows
that  the creation and annihilation operators satisfy the nilpotency
relations $(a^-)^{2s+1}= (a^+)^{2s+1}=0$. This means that the
corresponding representation is $(2s+1)$-dimensional. It is
interesting to note that
by using (\ref{sum}), one can write the condition (\ref{condition}) as 
\begin{eqnarray}
{\rm Tr}~G = 0.
\end{eqnarray}
where the trace is over the $(2s+1)$-dimensional Fock space. This
new algebra is covering the following results:
\begin{itemize}
\item (i)- The truncated harmonic oscillator introduced by Pegg-Barnett \cite{Pegg}: 
\begin{eqnarray}
 F(N) = N, \qquad G(N) = {\bf I} - (2s+1)\vert 2s \rangle \langle 2s \vert.
\end{eqnarray}
\item (ii)- The finite dimensional oscillator algebra ${\cal A_{\kappa}}$ ($\kappa=-1/2s < 0$) defined in \cite{Daoud}: 
\begin{eqnarray}
 F(N) = N \left[1 + \kappa(N-1)\right], \qquad G(N) = {\bf I} + 2\kappa N. 
\end{eqnarray}
\item (iii)- The truncated generalized oscillator algebra ${\cal A_{\kappa}}$ ($\kappa > 0$) \cite{Daoud}: 
\begin{eqnarray}
  F(N) = N \left[1 + \kappa(N-1)\right], \qquad  G(N) = {\bf I} + 2\kappa N - F(2s+1)\vert 2s \rangle \langle 2s \vert.
\end{eqnarray}
\end{itemize}
The eigenvalues of the operators $G(N)$, corresponding to the
special cases (i), (ii) and (iii) are respectively given by $G(n) =
1 - (2s+1)\delta_{n,2s}$, $G(n) = 1 + 2 \kappa n$ and $G(n) = 1 + 2
\kappa n - F(2s+1)\delta_{n,2s}$ for $n = 0, 1, \cdots, 2s$.
 At this
level, we have different comments in order. It is easily seen that
for $\kappa = 0$, the case (iii) reduces to (i). As mentioned
before, the algebra (\ref{wha}) is more general and covers many
other variants of generalized harmonic oscillators. The particular
cases (i), (ii) and (iii) are interesting in the context of quantum
optics. Indeed, the case (i) was introduced by Pegg and Barnett
\cite{Pegg} to define, in a consistent way, the phase operator. The
others cases (ii) and (iii) constitute an extension of the ideas
developed in~\cite{Pegg} to define phase operator for systems with
nonlinear spectrum \cite{Daoud}.

Using the algebraic structure of the generalized oscillator algebra
(\ref{wha}), one can introduce  an operator that generalizes the
Hamiltonian $a^+a^-$ for the one-dimensional harmonic oscillator.
Starting from
\begin{eqnarray}
a^+a^- \vert n \rangle  = F(n) \vert n \rangle
\end{eqnarray}
it is easy to realize the required Hamiltonian as
\begin{eqnarray}
H(N) \equiv F(N) = a^+a^-.
\end{eqnarray}
where $H(N)$ can be regarded as the Hamiltonian corresponding to
the algebra (\ref{wha}).
As usual, one can proceed with the eigenvalue equation
\begin{eqnarray}
H(N) \vert n \rangle  = F(n) \vert n \rangle
\end{eqnarray}
to get the solutions of the energy spectrum   of a quantum dynamical system described by 
$F(N)$. As illustration of the realized Hamiltonian, we give the
explicit forms for different cases mentioned above.
Indeed, we have for case (i)
\begin{eqnarray}
H(N)   = \sum_{n=0}^{2s} n \vert n \rangle \langle n \vert
\end{eqnarray}
and case (ii)
\begin{eqnarray}
H(N)   = \sum_{n=0}^{2s} \frac{n}{2s}( 2s + 1 - n) \vert n \rangle
\langle n \vert
\end{eqnarray}
as well as case (iii)
\begin{eqnarray}
H(N)  = \sum_{n=0}^{2s} n\left[1 + \kappa(n-1)\right] \vert n
\rangle \langle n \vert.
\end{eqnarray}

\section{ Temporally phase states}

We show how to build the phase states, which are temporally stable
under time evolution. For this, recall that our Hilbert space ${\cal
F}$ has $(2s+1)$-dimensions where the actions of the operators $a^-$
and $a^+$ on ${\cal F}$ are given in (\ref{action sur les n}). These
are supplemented by the condition
      \begin{eqnarray}
a^+ \vert 2s \rangle = 0
      \end{eqnarray}
which can easily be deduced from the calculation of expectation
value $\langle 2s \vert a^- a^+ \vert 2s \rangle$.

As $F(N)$ is a positive definite operator on the finite dimensional
Fock space, let us consider the  decomposition of the annihilation
$a^-$ and creation $a^+$  operators given by \cite{Daoud}
      \begin{eqnarray}
a^- = E~ \sqrt{F(N)}, \qquad  a^+ = \sqrt{F(N)}~ E ^{\dagger}
      \label{decompo cas fini}
      \end{eqnarray}
and one can show that the operator $E$ satisfies
      \begin{eqnarray}
E \vert n \rangle = e^{i [F(n) - F(n-1)] \varphi } \vert n-1
\rangle, \qquad n = 1, 2, \cdots, 2s
      \label{action de E}
      \end{eqnarray}
      which  is valid modulo $2s+1$. For
 $n=0$, we have 
            \begin{eqnarray}
E \vert 0 \rangle = e^{i [F(0)- F(2s)] \varphi} \vert 2s \rangle.
            \end{eqnarray}
It follows that for $E^{\dagger}$, one can obtain
      \begin{eqnarray}
E^{\dagger} \vert n \rangle = e^{-i [F(n+1) - F(n)] \varphi } \vert
n+1 \rangle
      \end{eqnarray}
where $n+1$ should be understood modulo $2s+1$.  Note that from above relations,  one can easily check that 
$E$ is an unitary operator. Therefore,  it is clear now that
(\ref{decompo cas fini}) constitutes a polar decomposition of the
operators $a^-$ and $a^+$.

To deduce 
the phase states associated with the finite dimensional algebra
defined in the previous subsection, let us in the beginning look at
the eigenstates of the operator $E$. In doing so, we solve the
eigenvalue equation
      \begin{eqnarray}
E \vert z \rangle = z \vert z \rangle, \qquad \vert z \rangle =
\sum_{n = 0}^{2s} C_n z^n \vert n \rangle
      \end{eqnarray}
where $z \in \mathbb{C}$. According to the method developed in
\cite{Daoud}, it is easy to see that the eigenvalues $z$ should
satisfy the  discretization condition
      \begin{eqnarray}
z^{2s+1} = 1
      \end{eqnarray}
and therefore the complex variable $z$ is a root of unity, such as
      \begin{eqnarray}
z = q^m \qquad m = 0, 1, \cdots, 2s
      \end{eqnarray}
where the $q$ parameter is defined as
\begin{eqnarray}
q := e^{2 \pi i / (2s+1)}. \label{definition of q}
\end{eqnarray}
Thus, these tell us that
the normalized eigenstates $\vert z \rangle \equiv \vert m , \varphi
\rangle$ of $E$ are of the form
\begin{eqnarray}
\vert m , \varphi \rangle = \frac{1}{\sqrt{2s+1}} \sum_{n=0}^{2s}
e^{-i F(n) \varphi} q^{mn} \vert n \rangle
\label{coherentstatemvarphi}
\end{eqnarray}
where the parameters $m \in \mathbb{Z}/(2s+1)\mathbb{Z}$ and
$\varphi \in \mathbb{R}$. It is easy to see that
\begin{eqnarray}
E \vert m , \varphi \rangle = e^{i\theta_m} \vert m , \varphi
\rangle, \qquad \theta_m = m \frac{2 \pi}{2s+1}
\label{operateurtheta}
\end{eqnarray}
which reflects that $E$ is indeed a phase operator and therefore
$\vert m , \varphi \rangle$ are the required phase states. Note
that, in the particular case $\varphi = 0$, the states $\vert m , 0
\rangle$ correspond to an ordinary discrete Fourier transform of the
basis $\{ \vert n \rangle : n = 0, 1, \cdots, 2s \}$ of the
$(2s+1)$-dimensional space ${\cal F}$. We show that the phase states
$\vert m , \varphi \rangle$ are temporally stable under ``time
evolution'', namely
\begin{eqnarray}
e^{-i H(N) t} \vert m , \varphi \rangle = \vert m , \varphi + t
\rangle
\end{eqnarray}
for any value of the real parameter $t$. The parameter $\varphi$
plays a key role to ensure the temporal stability of the states
$\vert m , \varphi \rangle$.
 For fixed $\varphi$, they satisfy the equiprobability relation
\begin{eqnarray}
| \langle n \vert m , \varphi \rangle | = \frac{1}{\sqrt{2s+1}},
\qquad n, m \in \mathbb{Z}/(2s+1)\mathbb{Z}. \label{computational et
MUB}
\end{eqnarray}
They constitute an  orthonormal set
      \begin{eqnarray}
\langle m , \varphi \vert m' , \varphi \rangle = \delta_{m,m'}
\qquad m, m' \in \mathbb{Z}/(2s+1)\mathbb{Z}
      \end{eqnarray}
for fixed $\varphi$ and satisfy the closure property
      \begin{eqnarray}
\sum_{m = 0}^{2s} \vert m , \varphi \rangle \langle m , \varphi
\vert = {\bf I}.
      \end{eqnarray}
Finally, it is interesting to note that the temporally stable phase
states are not all orthogonal. Indeed, the overlap between two phase
states $\vert m' , \varphi' \rangle$ and $\vert m , \varphi \rangle$
reads as
   \begin{eqnarray}
\langle m , \varphi \vert m' , \varphi' ) = \frac{1}{2s+1}
\sum_{n=0}^{2s} q^{\rho(m-m', \varphi - \varphi', n)}
   \label{overlap}
   \end{eqnarray}
where the function $\rho$ is
         \begin{eqnarray}
\rho(m-m', \varphi - \varphi', n) = - (m - m')n + \frac{2s+1}{2\pi}
(\varphi - \varphi') F(n)
         \end{eqnarray}
and $q$ is given in (\ref{definition of q}).

 At this level, we would like to emphasize the key role of the phase
parameter $\varphi$ introduced starting from relations (\ref{action sur les n}) 
that defines the actions of creation and annihilation operators. 
Indeed, firstly $\varphi$ ensures the temporal stability of the
phase states $\vert m , \varphi \rangle$
(\ref{coherentstatemvarphi})  under time evolution. Secondly, if one
ignores $\varphi$, i.e. $\varphi = 0$, $\vert m , \varphi \rangle$
reduce to those derived by Begg and Barnett using the truncated
harmonic oscillator \cite{Pegg}. This means
that for two 
generalized Weyl--Heisenberg algebras characterized by different
structure functions $F(N)$, we end up with the same phase states.
Thus, to differentiate between these states, the presence of
$\varphi$ is necessary in their forms. In this sense, it must be
noticed that the $SU(2)$ phase states obtained in \cite{Vourdas} are
similar
to those 
of Pegg and Barnett despite the fact that the involved symmetries
are different. Hence, to avoid such  problem we introduced
$\varphi$ that
 allows us to keep the trace of the symmetry and dynamics of the system under consideration. Note
also that  $\varphi$ plays a key role in relating phase states and
mutually unbiased bases \cite{Daoud}.

\section{ Beam splitting of phase states and entanglement}

We briefly describe the effect of beam splitting on a known input
state with the vacuum at the second port.
We assume that the vertical input beam is always prepared in the
Fock ground state
and the  state of interest is the input state in the horizontal
beam, which by convention precedes that of vertical beam.
Algebraically, the beam splitting transformation is described as
follows. The input field described by the usual harmonic oscillator
operator $a_1^{\pm}$ is superposed on the other input field with
operator $a_2^{\pm}$ by a lossless symmetric beam splitter, with 
the  transmission $t$ and reflection $r$ coefficients. 
The output field operators $a_3^{\pm}$ and $a_4^{\pm}$ are given by
\begin{equation}
a_3^{\pm} =  {\cal B}(\theta) a_1^{\pm}  {\cal B}^{\dagger}(\theta),
\qquad a_4^{\pm} =  {\cal B}(\theta) a_2^{\pm}  {\cal
B}^{\dagger}(\theta)
\end{equation}
where the unitary beam splitter operator of angle $\theta$ is
\begin{equation}
 {\cal B}(\theta) = \exp\left[\frac{i}{2}\theta \left(a_1^+a_{2}^- + a_1^-a_{2}^+\right)\right].
\end{equation}
The action of  ${\cal B}(\theta)$ on the state $ \vert n , 0
\rangle$ reads as
\begin{equation}
 {\cal B}(\theta) \vert n , 0 \rangle  =  \sum_{p=0}^{n} {\sqrt \frac{n!}{p!(n-p)!}} t^p (ir)^{n-p}\vert p, n-p \rangle
\end{equation}
where the transmission and reflection coefficients are defined by
\begin{equation}
 t = \cos\frac{\theta}{2}, \qquad  r = \sin\frac{\theta}{2}.
\end{equation}

It is well known  that among pure states of a single mode field,
only harmonic oscillator or Glauber coherent states do not become
entangled upon beam splitting. Indeed, for the 50:50 beam splitter
 with an ordinary Glauber coherent
state incident on one port and a vacuum  on the
other, the beam splitting result is a product of two Glauber states. This procedure do not provides us with 
 an entangled system.
In general, any other pure state at the input results in an
entangled state at the output. This explains how the beam splitter
acts as entangler and why it is used in experiments as device to
generate entangled states.

Now, let us proceed in our case to rewrite the action of the beam
splitter operator on the phase states (\ref{coherentstatemvarphi})
as
\begin{equation}
 {\cal B}(\theta) \vert m,\varphi\rangle \otimes \vert 0 \rangle = \frac{1}{\sqrt{2s+1}} \sum_{n=0}^{2s} \sum_{p=0}^{n}
\sqrt{\frac{n!}{(n-p)!p!}}~q^{mn} t^p (ir)^{n-p}~ e^{-iF(n)\varphi}~
\vert p , n-p \rangle.
\end{equation}
The corresponding output density is then given by
\begin{equation}
\rho_{1,2} =   {\cal B}(\theta) \vert m,\varphi \rangle \otimes
\vert 0 \rangle \langle m,\varphi \vert \otimes \langle 0 \vert
{\cal B}^{\dagger}(\theta)
\end{equation}
It follows that one can write the bipartite reduced density matrix
$\rho_{12}$, which is obtained by tracing out the second system.
This is
\begin{equation}
\rho_1 =  {\rm Tr}_{2} \rho_{1,2}
\end{equation}
where label $2$  is reflecting the trace on second states. It is
easy to see that the reduced density matrix is given by
\begin{equation}
\rho_{1} = \sum_{n=0}^{2s} \sum_{n'=0}^{2s} \sum_{l=0}^{{\rm
min}(2s-n, 2s-n')} c(n,l)\overline{c (n',l)} \vert n\rangle \langle
n' \vert
\end{equation}
where the coefficients  $c(n,l)$ have the form
\begin{equation}
 c(n,l) = \frac{1}{\sqrt{2s+1}}
\sqrt{\frac{(n+l)!}{n!l!}}~q^{m(n+l)} t^n (ir)^{l}~
e^{-iF(n+l)\varphi}.
\end{equation}


Next, we are interested in the amount of entanglement of the beam splitter output states. 
In doing so, we
examine the entanglement of phase states when passed through a beam
splitter
by using the linear entropy  \cite{Bose} 
\begin{equation}
S = 1 - {\rm Tr}(\rho_1^2). \label{entropy}
\end{equation}
Clearly, it goes to zero for a pure state. 
After some algebra, 
we show that (\ref{entropy})
rewrites 
\begin{equation}
S = 1 - \sum_{n=0}^{2s} \sum_{n'=0}^{2s} \sum_{l=0}^{{\rm min}(2s-n,
2s-n')}  \sum_{l'=0}^{{\rm min}(2s-n, 2s-n')} s(n,n',l,l')
\label{linearentropy}
\end{equation}
where $s(n,n',l,l')$ is given by
\begin{equation}
s(n,n',l,l') = \frac{1}{(2s+1)^2} e^{-i \phi(n,n',l,l')}
\sqrt{(n+l)!(n'+l')!(n+l')!(n'+l)!} \frac{t^{2(n+n')}}{n!n'!}~
\frac{r^{2(l+l')}}{l!l'!}
\end{equation}
and the phase term reads as
\begin{equation}
\phi(n,n',l,l') =  \left[F(n+l)+F(n'+l')-
F(n'+l)-F(n+l')\right]\varphi.
\end{equation}
We have some remarks in order. Firstly, according to last equations we notice that 
the linear entropy is $m$-independent. Secondly, one can verify that
the functions $\phi(n,n',l,l')$ satisfy some symmetries with respect
to interchange of their quantum numbers. These are
\begin{equation}
\phi(n,n',l,l') = - \phi(n,n',l',l), \qquad \phi(n,n',l,l') =
\phi(n',n,l,l')
\end{equation}
which can be used to express  the linear entropy in terms of
$\cos\phi(n,n',l,l')$ instead of $e^{-i \phi(n,n',l,l')}$ together
with the symmetry of the summations in (\ref{linearentropy}).

For the temporally stable phase states considered in section 3, we
study numerically the behavior of the first-order entropy as a
function of the relevant parameters $s$, $\varphi$ and the
reflection coefficient $r$ as well. We shall focus on the phase
states associated with the generalized Weyl--Heisenberg algebra
defined through the structure function
\begin{equation}
F(N) = \frac{N}{2s}\left(2s+1-N\right).
\end{equation}
We first plot the entanglement against $\varphi$ and $r^2$ for
different finite dimensional Hilbert of $(2s+1)$-dimensions.
\begin{center}
  \includegraphics[width=5in]{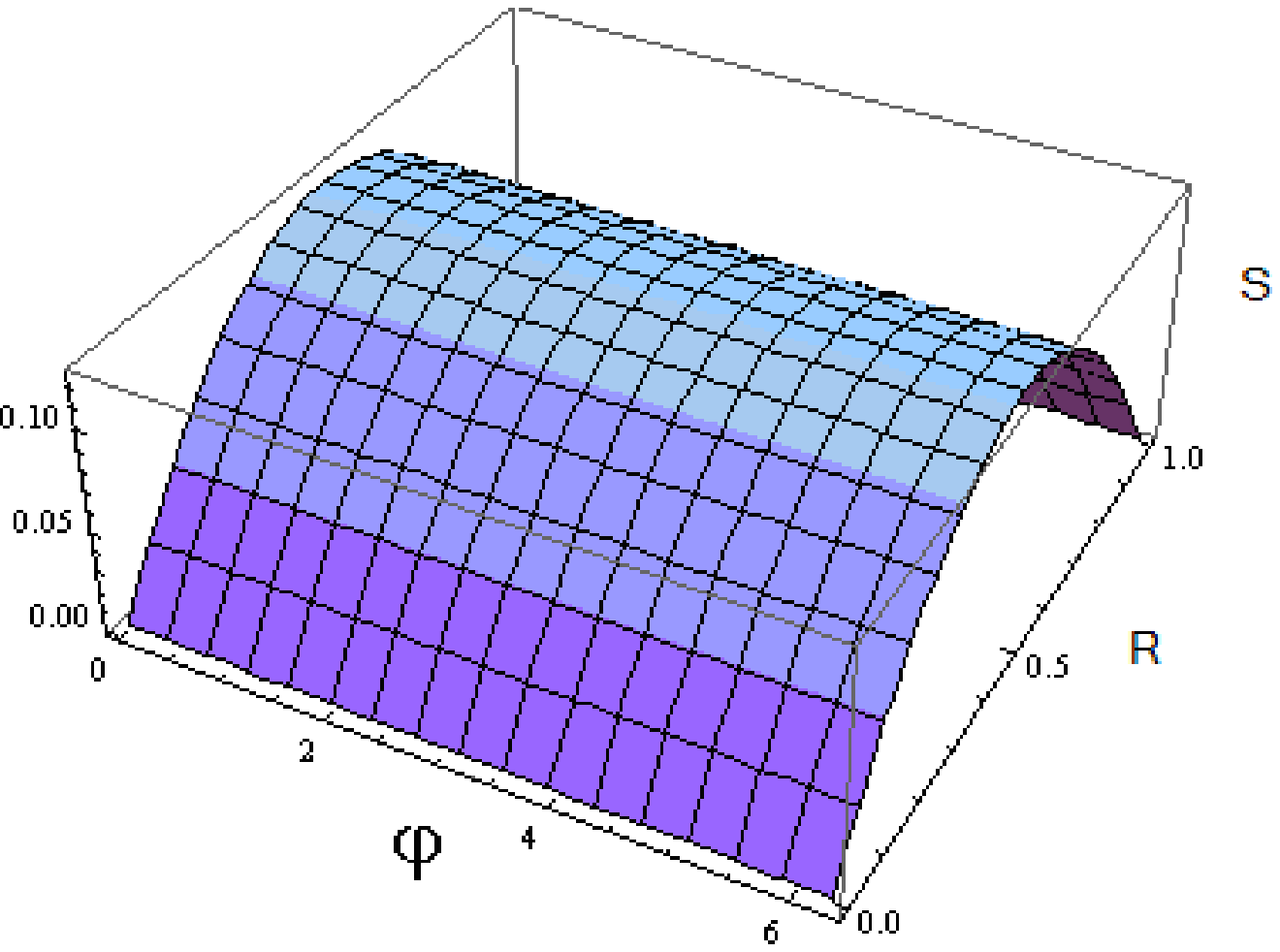}\\
{\sf Figure 1:  {Linear entropy of phase states as a function of $R
= r^2$ and $\varphi$ for $s = 1/2$.}}
\end{center}
\begin{center}
  \includegraphics[width=5in]{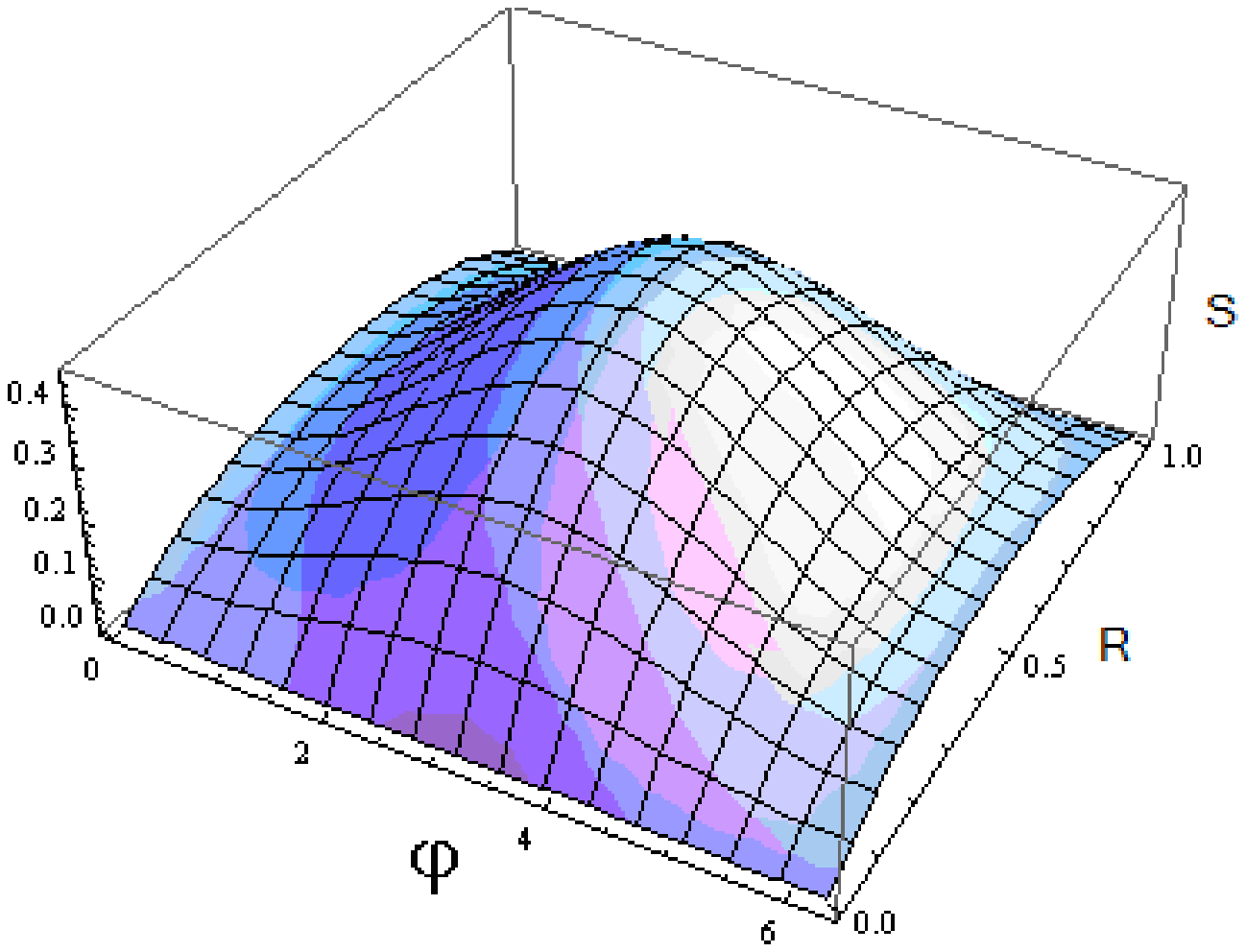}\\
{\sf Figure 2:  {Linear entropy of phase states as a function of $R
= r^2$ and $\varphi$ for $s = 1$.}}
\end{center}
\begin{center}
  \includegraphics[width=5in]{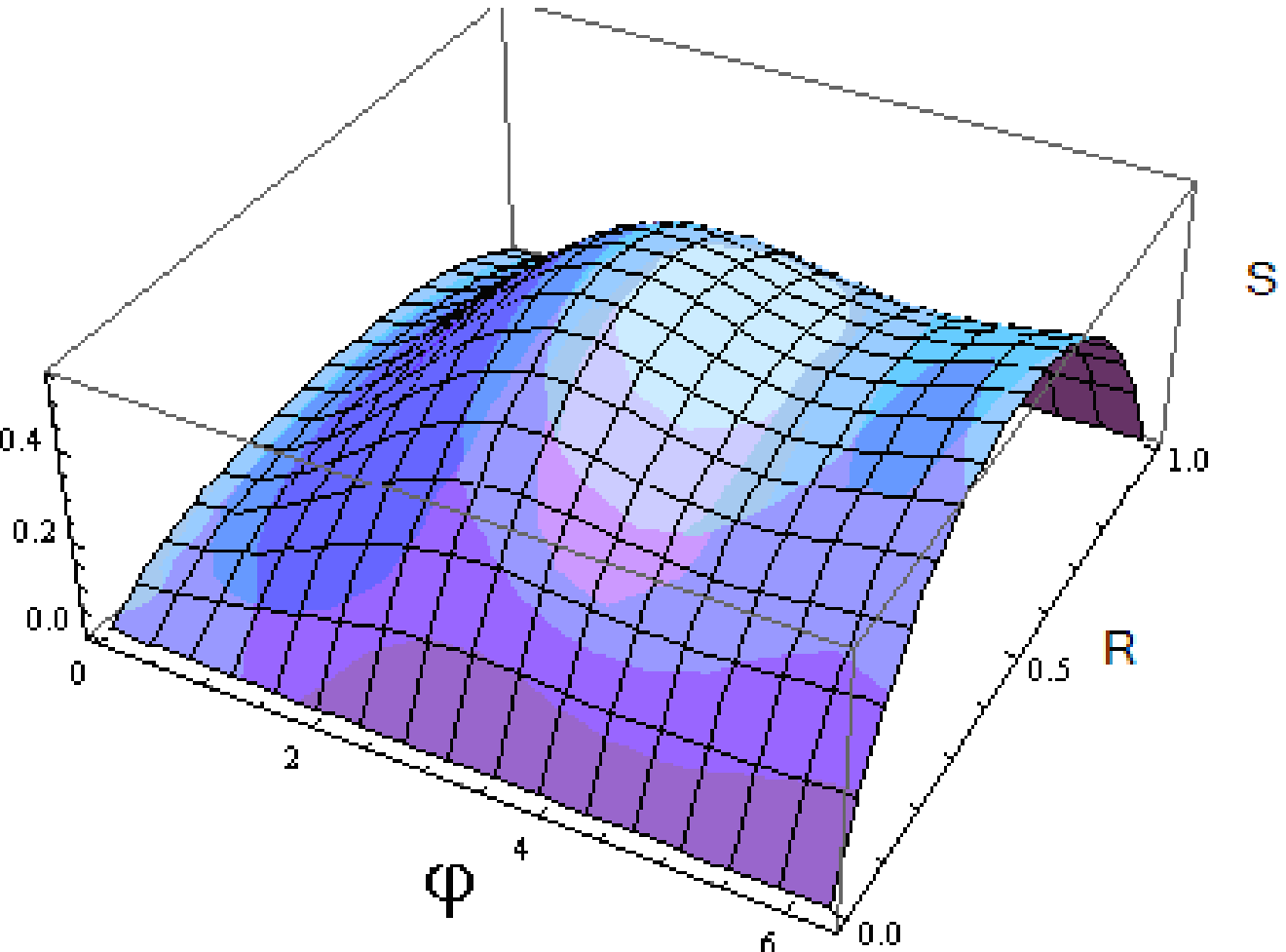}\\
{\sf Figure 3:  {Linear entropy of phase states as a function of $R
= r^2$ and $\varphi$ for $s = 3/2$.}}
\end{center}
According to the above figures, we note that for all value taking by
$s$ there is a maximum entanglement at the point $r^2 = 1/2$, which
corresponds to a 50:50 beam splitter. Furthermore, we have  two
interesting features in order. The first is that for qubits $ s=
1/2$, the degree of entanglement is $\varphi$-independent as shown
in Figure 1. However, for qutrits $( s = 1)$ and fixed reflection
coefficient, the behavior of the linear entropy is Gaussian. The
maximum is reached for $r^2 = 1/2$ and $\varphi = \pi$, see Figure
2. This changes completely for higher dimensional Hilbert spaces.
Indeed, for $s = 3/2$, for fixed value of $r^2$, the linear entropy
undergoes an initial rapid increase  to reach the maximum entropy
for $\varphi = \pi$ followed by a slower drop and then increases
when $\varphi$ does go to $2\pi$ as it shown in Figure 3.

As we explained above, the parameter $\varphi$ plays a key role in
defining the phase states and ensuring the temporal stability of
phase states. It follows that it is interesting to examine the
linear entropy in term of  $\varphi$ for $s$ fixed and $r^2 = 1/2$
for which one has the maximum of entanglement of phase states.
\begin{center}
  \includegraphics[width=4in]{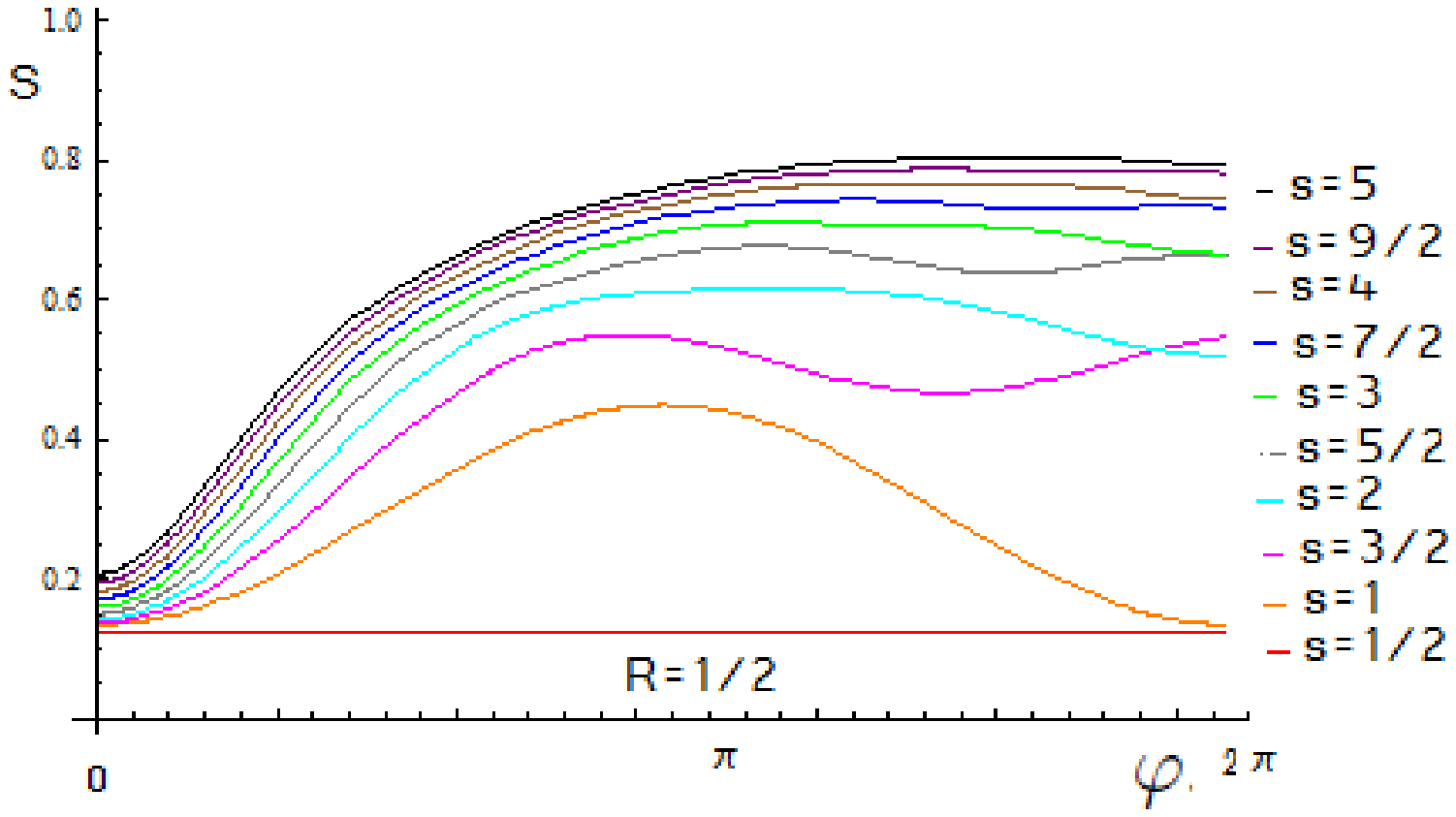}\\
{\sf Figure 4:  {Linear entropy of phase states after being passed
through a 50:50 beam splitter against $\varphi$ for $s = 1/2, 1,
\cdots, 5$.}}
\end{center}
The behavior of linear entropy as function of the parameter
$\varphi$ is represented in Figure 4. It is easily seen that for
$s=1/2$, the linear entropy is constant. For qutrits, i.e. $s =1$,
the entropy looks like a gaussian with a maximum around $\varphi =
2\pi$. For higher levels quantum systems, the linear entropy behaves
differently. For instance, with phase states of quartits, i.e. $s =
3/2$, the degree of entanglement increases for $\varphi < \pi$,
decreases for $\pi < \varphi < 3\pi/2$ and increases after. As the
parameter $\varphi$ is deeply related to time evolution of phases
states, the figure 4 can be viewed as representing the temporal
evolution of the degree of entanglement of phase states.

In Figure 5 we plot  the linear entropy  of  phase states versus the
parameter $s$ for different values of $\varphi$ to underline its
basic behavior.
\begin{center}
  \includegraphics[width=4in]{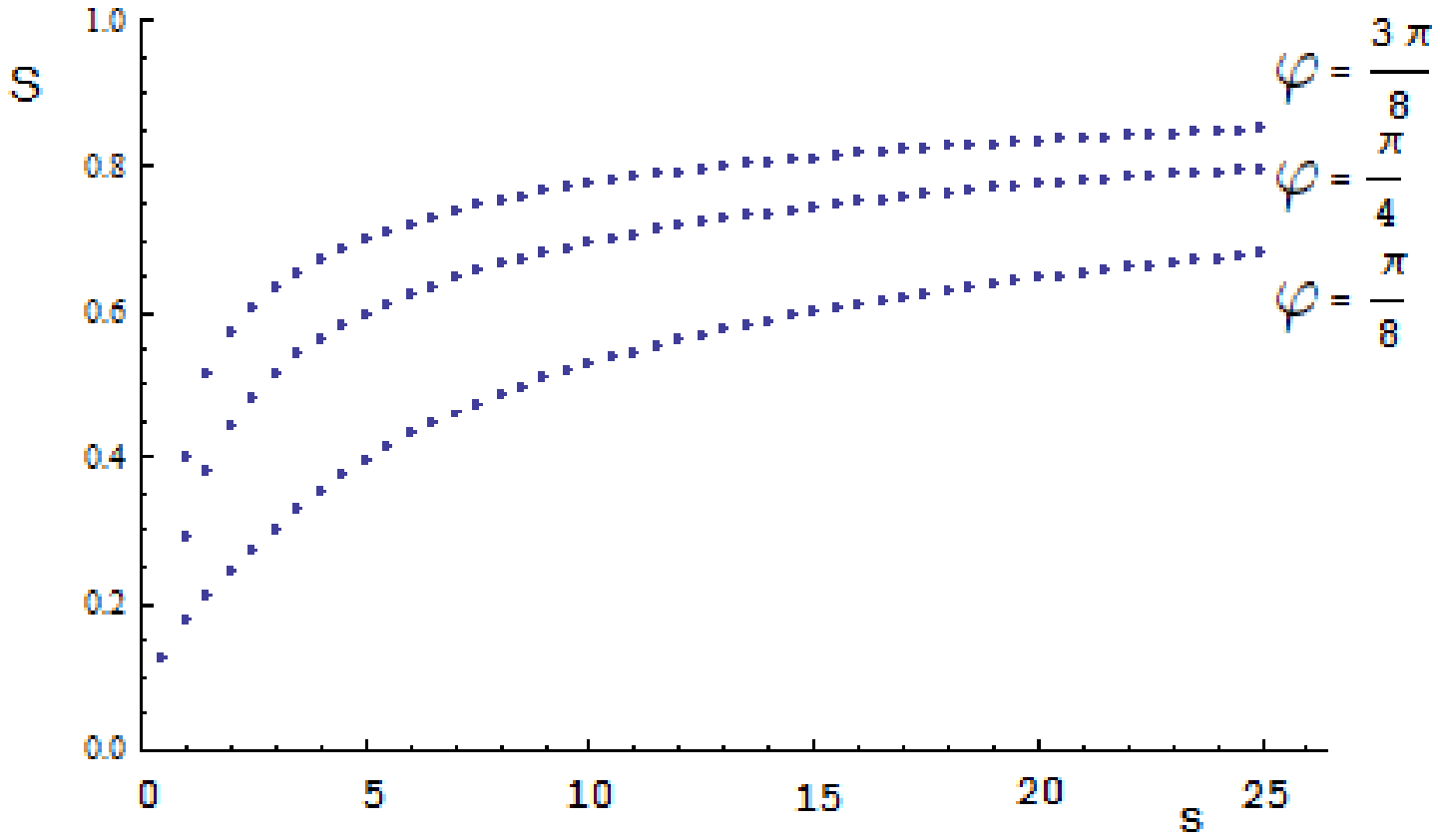}\\
{\sf Figure 5:  {Linear entropy of phase states after being passed
through a 50:50 beam splitter against $s$ for different values of
$\varphi$.}}
\end{center}
Following this figure, we notice that the linear entropy initially
rapidly rises for lower $s$ but increases slowly for $s \geq 10 $.
However for higher $s$, the linear entropy does not approaches zero.
This shows that the degree of entanglement is increasing when the
dimension of the generalized Weyl--Heisenberg becomes large.

\section{\bf Concluding remarks}

The main idea of the present work is the investigation of the degree
of entanglement of temporally phase states. The latter are
constructed as eigenstates of a unitary phase operator resulting
from the polar decomposition of ladder operators of finite
dimensional Weyl--Heisenberg algebra. Using the beam splitter as
entangler, we computed the degree of entanglement of phase states by
mean of linear entropy. We investigated the behavior of this
quantity as function of the dimension of the Hilbert space, the
transmission-reflection coefficients of the beam splitter and the
parameter ensuring the temporal stability of the phase states.

It is clearly shown that the maximal entanglement is provided by
50:50 beam splitter. Also, the entanglement increases with
increasing dimension of the system. For instance the phase states
for qutrits are more entangled than qubits. The entanglement of
phase states when passed through a beam splitter is strongly
dependent on the parameter $\varphi$ which play an essential role in
defining the eigenstates of the unitary phase operator and ensure
their temporal stability.

Finally, we note that we did not consider the limiting case $s
\longrightarrow \infty$. In fact for infinite dimensional Hilbert
space, one can not define unitary phase operator in a consistent
way. Consequently the phase states associated with the ordinary
Weyl--Heisenberg can not be constructed using the formalism
developed in this work.

\section*{Acknowledgments}
MD would like to express his thanks to Max Planck Institute for
Physics of Complex Systems (Dresden-Germany) where this work was
done. The authors would like to thank A. Jellal for his careful
reading of the manuscript and for his valuable comments.


\end{document}